\documentclass[preprint,prb,a4paper,showabstract,superscriptaddress ]{revtex4-2}
\usepackage{multirow}
\usepackage{bm}

\usepackage{graphicx}% Include figure files
\usepackage{epstopdf}
\usepackage{dcolumn}% Align table columns on decimal point
\usepackage{bm}% bold math
\usepackage{hyperref}% add hypertext capabilities
\usepackage[utf8]{inputenc} % allow utf-8 input
\usepackage[T1]{fontenc}    % use 8-bit T1 fonts
\usepackage{lmodern}
\usepackage{url}            % simple URL typesetting
\usepackage{booktabs}       % professional-quality tables
\usepackage{amsfonts}       % blackboard math symbols
\usepackage{nicefrac}       % compact symbols for 1/2, etc.
\usepackage{microtype}      % microtypography
\usepackage[version=3]{mhchem}
\usepackage{natbib}
\usepackage{amsmath}
\usepackage{amssymb}
\usepackage{xcolor}
\usepackage[range-units = single,range-phrase = \text{--},separate-uncertainty = true,detect-all]{siunitx}
	\DeclareSIUnit\linepair{lp}
	\DeclareSIUnit\pixels{px}
\usepackage{soul}
\setstcolor{red}
\usepackage{cancel}

\definecolor{bleudefrance}{rgb}{0.19, 0.55, 0.91}

\newcommand{\R}{\mathbf{R}}

\begin{document}

%\title{Passive Retrieval of the Transmission Matrix\\ for Deep Imaging of a Multiple Scattering Medium}
%\title{Non-invasive Retrieval of the Transmission Matrix  Deep Inside a Multiple Scattering Medium for 3D Imaging with Incoherent Light}
%\title{Deep adaptive focusing inside scattering media with incoherent light}
%\title{Deep passive focusing inside multiple scattering media}
%\title{In-depth transmission matrix in multiple scattering media}
%\title{In-depth transmission of light inside multiple scattering media}
%\title{In-depth focusing of incoherent light inside multiple scattering media}
%\title{In-depth memory effect for incoherent light in multiple scattering media}
\title{Self-Portrait of the Focusing Process in Speckle: \\ III. Tailoring Complex Spatio-Temporal Focusing Laws To Overcome Reverberations in Reflection Imaging}

\author{Elsa Giraudat}
\affiliation{Institut Langevin, ESPCI Paris, PSL University, CNRS, 75005 Paris, France}
\author{Flavien Bureau}
\affiliation{Institut Langevin, ESPCI Paris, PSL University, CNRS, 75005 Paris, France}
\author{William Lambert}
\affiliation{SuperSonic Imagine, Aix-en-Provence, France}
\author{Mathias~Fink}
\affiliation{Institut Langevin, ESPCI Paris, PSL University, CNRS, 75005 Paris, France}
\author{Alexandre Aubry$^*$}
\affiliation{Institut Langevin, ESPCI Paris, PSL University, CNRS, 75005 Paris, France}

\date{\today}
\begin{abstract}
    \textbf{This is the third article in a series of three dealing with the exploitation of speckle for imaging purposes. In complex media, a fundamental limit is the multiple scattering phenomenon that completely blurs the imaging process in depth. Matrix imaging can provide a relevant framework for solving this problem. As it proved to be an adequate tool for probing reverberations in speckle [E. Giraudat et al., Part I], we will show how it can be used to tailor complex spatio-temporal focusing laws to monitor the interference between the multiply-reflected paths and the ballistic component of the wave-field. To do so, we extend the distortion matrix concept to the frequency domain. An iterative phase reversal process operated from the space-time Fourier space is then used to compensate for reverberations and optimize both the axial and transverse resolution of the confocal image. Here, we first present an experimental proof-of-concept consisting in imaging a tissue-mimicking phantom through a reverberating plate before outlining the potential and the limits of this strategy for transcranial ultrasound and beyond.}
\end{abstract}
%\keywords{imaging, microscopy, multiple scattering, transmission matrices, reflection matrices}
\maketitle

Multiple scattering of waves in heterogeneous media is a fundamental problem with important applications, ranging from seismic wave propagation through the Earth's crust to deep tissue imaging in microscopy or trans-cranial ultrasound for brain imaging. Conventional focusing and imaging techniques based on Born approximation generally fail in strongly scattering media because of the multiple scattering  events undergone by the incident wavefront. To compensate for those wave-front distortions induced by the medium heterogeneities, numerous adaptive focusing methods have been developed~\cite{Booth2014,ali_aberration_2023}. These approaches generally consist in an application of time delays or phase shifts to each transmission and reception channel of the imaging device. While they are efficient, to a certain extent, to compensate for aberrations induced by forward multiple scattering trajectories~\cite{kang_tracing_2023,najar_non-invasive_2023}, they cannot overcome the problem of reverberations that require the tailoring  of more complex spatio-temporal focusing laws. Each frequency component of the back-scattered echoes shall be actually compensated independently in order to make multiply-scattered waves associated with a broad time-of-flight distribution interfere constructively at any point inside the medium~\cite{Mosk2012}. 

\begin{figure}[h!tb]\centering
\includegraphics{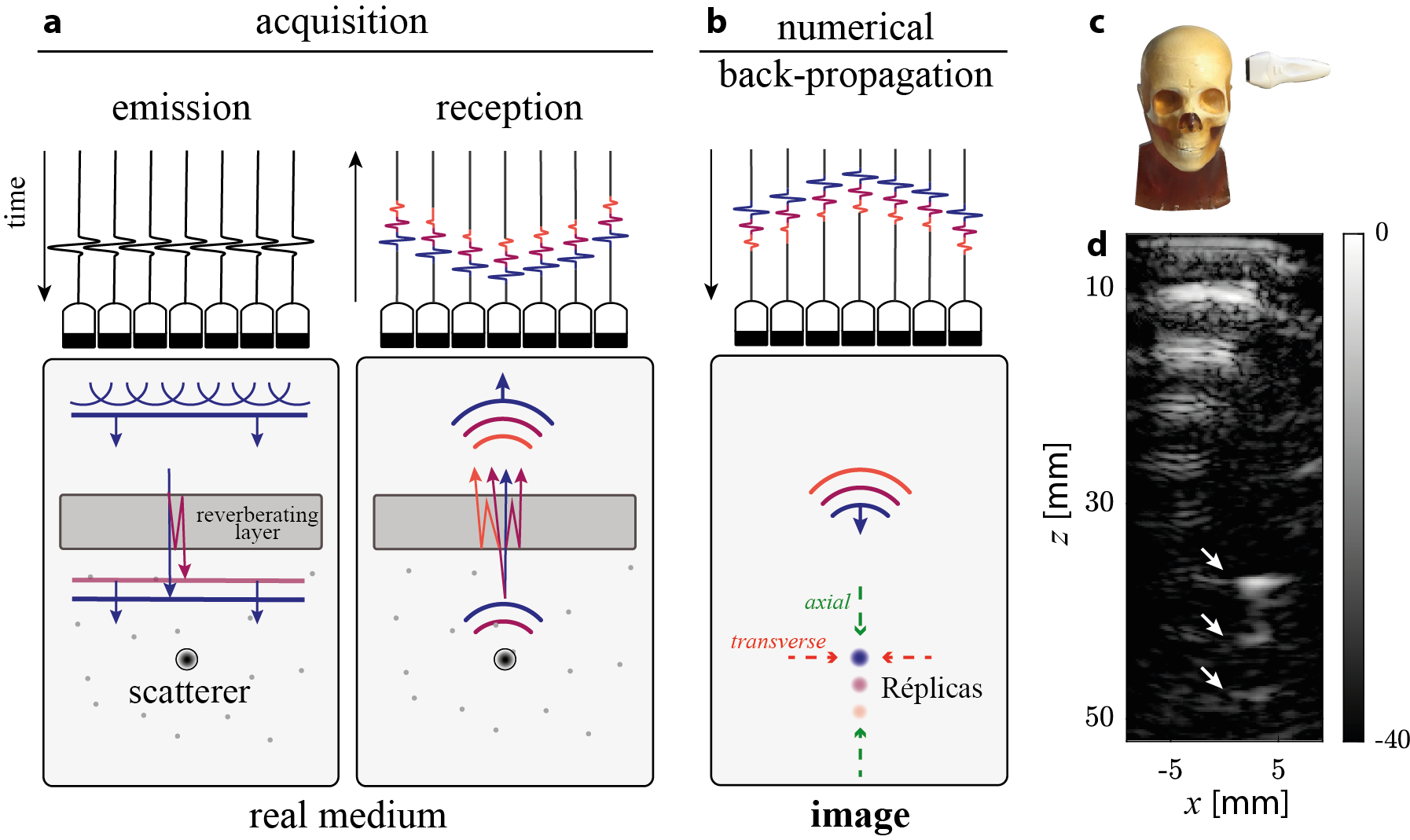}
    \caption{\textbf{Artifacts caused by reverberations in ultrasound}.  \textbf{a} A medium is insonified by a plane wave. After passing through a reverberating layer, the plane wave is split into a tail of multiply-reflected waves.  \textbf{b} The wave reflected by a strong diffuser in the medium produces a set of echoes, that are multiplied again when they pass back through the reverberating layer.  \textbf{c} During the image reconstruction stage using a homogeneous velocity model, the multiple reflected echoes, associated with longer times-of-flight, are back-propagated to overestimated depths, giving rise to ghost images, or replicas, of the strong scatterer. Only a dispersive focusing law could correct for these artifacts.  \textbf{d}, Example of a confocal image obtained on an ultrasound head phantom containing a bright target, appearing tripled due to reverberations induced by the skull.}
    \label{ch4_RMintro}
 \end{figure}

In this paper, we focus on the problem of reverberations in ultrasound which is particularly prominent in transcranial imaging. Standard ultrasound actually relies on a focused beamforming process in which only the ballistic component of the wave-field is used to construct an image of the medium reflectivity (Fig.~\ref{ch4_RMintro}). This shortcoming can be the source of multiple artifacts in complex media. Figure~\ref{ch4_RMintro}d illustrates this last statement by displaying the ultrasound image obtained on a head phantom reproducing the acoustic characteristics of an adult skull and containing a spherical target for evaluating the aberrations induced by the skull layer. The reverberations generated at the skull-tissue interfaces induce multiple ghost images of the bright scatterer. Furthermore, although they are not central to our discussion below, it should also be noted that the multiple specular reflections at the skull interfaces also seriously degrade image quality at shallow depths. 

In the past, the problem of ultrasound focusing through the skull has been addressed in different ways. On the one hand, a first strategy consists in modeling the skull to estimate the time delays that need to be applied in order to correctly focus ultrasound. However, this requires prior knowledge of the geometry and acoustic properties of the skull, as well as the ability to predict wave propagation in this complex medium. The characteristics of the skull can be obtained in several ways, whether it be with X-rays~\cite{aubry_experimental_2003,pichardo_multi-frequency_2011,vyasPredictingVariationSubject2016,leungRapidBeamSimulation2019,webbAcousticAttenuationMultifrequency2021}, MRI~\cite{sunFocusingTherapeuticUltrasound1998, hynynenDemonstrationPotentialNoninvasive1998,suTranscranialMRImaging2020,webbAcousticAttenuationMultifrequency2021,leungComparisonMRCT2022} or using ultrasound data itself \cite{wydraDevelopmentPracticalUltrasonic2013,hajianAccurate3DProfile2017,nguyenminhEstimationThicknessSpeed2020,renaudMeasuringAnisotropyElastic2020,duNovelTranscranialUltrasound2020}. Coupling different imaging modalities can be quite tedious, as the success of the correction depends on the correct positioning of the probe. Moreover, these methods have been gradually refined to take into account refraction \cite{lindseyRefractionCorrection3D2014,shapooriUltrasonicAdaptiveBeamformingMethod2015, jiangRayTheoryBasedTranscranial2019,mozaffarzadeh_refraction-corrected_2022} and shear waves \cite{clement_enhanced_2004}. The more the approach is sophisticated, the more it is sensitive: a small error in estimating the thickness or mechanical properties of the skull can lead to a significant error in estimating the delay laws. Last but not least, a significant computation time is required for methods involving simulations~\cite{wang_transcranial_2013,mozaffarzadeh_lamb_2021}, which constitutes a barrier to the practical use of these methods \cite{aubryBenchmarkProblemsTranscranial2022,anglaTranscranialUltrasoundSimulations2023}.

An alternative strategy has been to employ adaptive focusing techniques inspired from astronomy. The broadband nature of ultrasound has led to the emergence of time reversal processing to properly harness the multiply-scattered waves through the skull. However, such approaches require either an active source inside the brain~\cite{thomasUltrasonicBeamFocusing1996,hynynenDemonstrationPotentialNoninvasive1998} such as a miniaturized transducer or the use of artificially induced ultrasonic stars such as a bubble that can be generated by cavitation~\cite{pernot_ultrasonic_2014} or vaporization \cite{haworth_towards_2008}. Such a process can then be coupled with beam steering techniques to scan the focal spot around the initial source location \cite{clementMicroreceiverGuidedTranscranial2002}. However, it relies on an angular memory effect that turns to be very limited through the skull, especially for multiply-reflected waves. It also means that an active or passive source shall be used for each isoplanatic volume inside the region of interest. The constraints of in vivo imaging are such that a non-invasive reflection correction method that does not require the presence of a guide star is required. 

To meet these criteria, it has been shown that spatial correlations of back-scattered echoes induced by a focused incident wave-front can be used to extract aberration laws \cite{montaldo_time_2011}.  To do so, an average over different incident focal spots is performed to synthesize a virtual guide star from different realizations of disorder. However, this experimental demonstration was restricted to the compensation of wave distortions and not to reverberations. Moreover, it rely on an hypothesis of local isoplanicity that implies a repetition of the whole process for each point in the field of view. It therefore requires a particularly time-consuming iterative focusing process that makes this approach impractical for trans-cranial imaging.
 
To avoid this tedious focusing process, the reflection matrix approach is particularly relevant~\cite{lambert_distortion_2020,bureau_three-dimensional_2023}. Once recorded, this matrix contains all the information available on the medium and any iterative focusing process can now be performed in post-processing in real time. In this paper, we will show how the time-focused reflection matrix and the frequency-dependent distortion matrix introduced in the first paper of the series~\cite{Giraudat2025} can be used to address each frequency component of the reflection matrix independently and thus go beyond aberration correction involving a simple application of time delays to recorded signals. In the first part, we propose to focus on the frequency degrees of freedom. A method for compensating temporal dispersion phenomena is presented and illustrated using the academic experiment involving an ultrasound phantom placed behind a reverberating plate~\cite{Giraudat2025}.  In the second part, the method is refined to address each angular component of the ultrasonic field independently. Finally, in the last part, this method is applied to the more ambitious case of a head phantom with a particularly heterogeneous skull bone. The potential and limits of matrix imaging for trans-cranial applications are established before showing the interest of the proposed method in other fields of wave physics such as optical microscopy and seismology. 
  
\begin{figure}[h!tb]\centering
    \includegraphics[width=\textwidth]{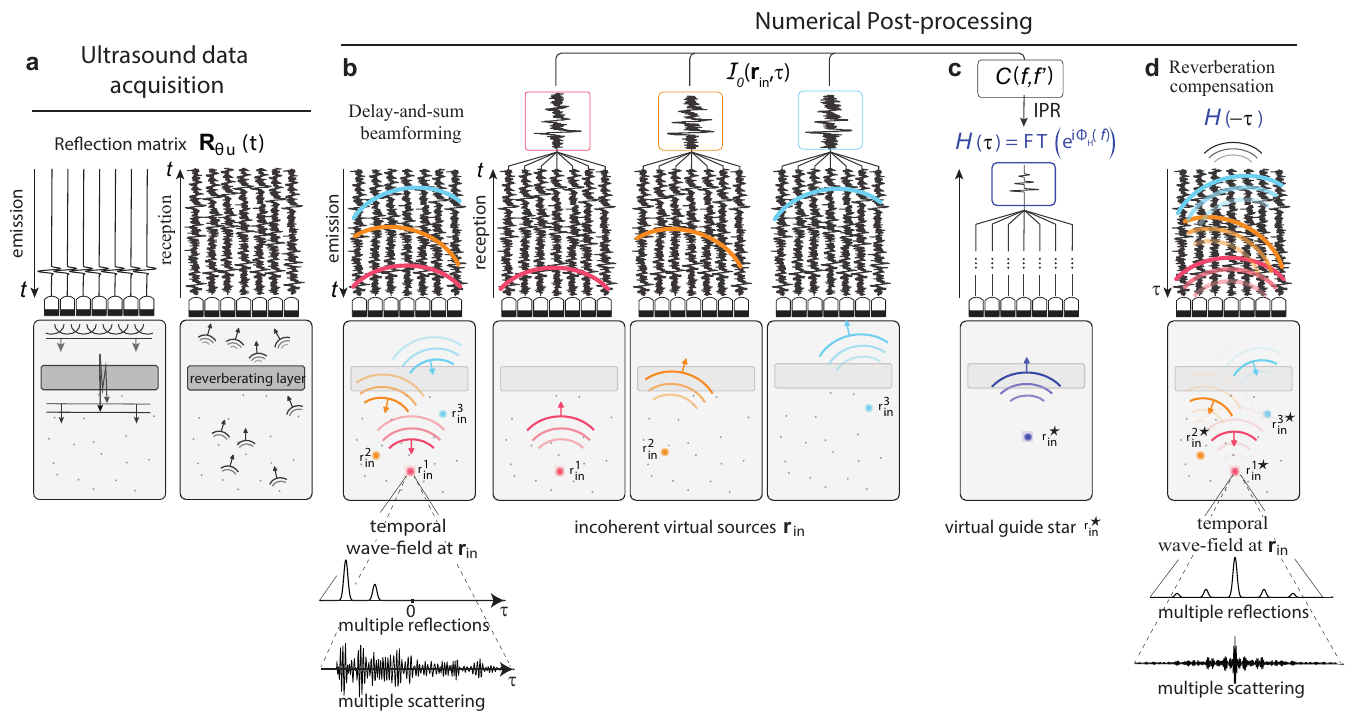}
    \caption{Extraction of a frequency dispersion law in speckle. \textbf{a} Acquisition of the reflection matrix $\mathbf{R}_{\mathbf{u}\bm{\theta}}(t)$: (\textit{i}) each incident plane wave is multiply-reflected by the reverberating layer before being scattered by sub-resolved heterogeneities; (\textit{ii}) the reflected waves also undergo reverberations before being recorded by the ultrasonic probe. \textbf{b} Confocal delay-and-sum beamforming process at each point $\mathbf{r}$ in the medium. The temporal dispersion of the echoes dispersed is captured by investigating the confocal signal $\mathcal{I}_0(\mathbf{r},\tau)$ as a function of time lapse $\tau$ with respect to the expected ballistic time. \textbf{c} An IPR analysis conducted on the correlation matrix between the backscattered echoes at each frequency leads to the synthesis of a coherent virtual source and the access to the associated reverberated signal $H(\tau)$ that accumulates the axial aberrations of the incident and reflected paths. \textbf{d} Time reversal of this signal yields a wave-front that can compensate for the axial aberrations of the reflection matrix, leading to a temporal recompression of the echoes associated with each point $\mathbf{r}$ of the medium.}
    \label{principle}
 \end{figure}

\section{Dispersive correction of reverberations}\label{ch4_section_CorrFreq}

In this work, we propose an approach to compensate for reverberations, particularly in the difficult case of random scattering media encountered in medical ultrasound in absence of any guide star. To begin with, we propose to focus on the frequency degrees of freedom of the wave-field. In the first paper of the series~\cite{Giraudat2025}, we saw that the temporal reflection matrix can be used to synthesize a coherent virtual reflector in speckle. Here, we will show that this virtual guide star can be used to tailor a dispersive focusing law in order to compensate for reverberations. 

\subsection{Proof-of-concept experiment}

\begin{figure}[h!tb]\centering
    \includegraphics[width=\textwidth]{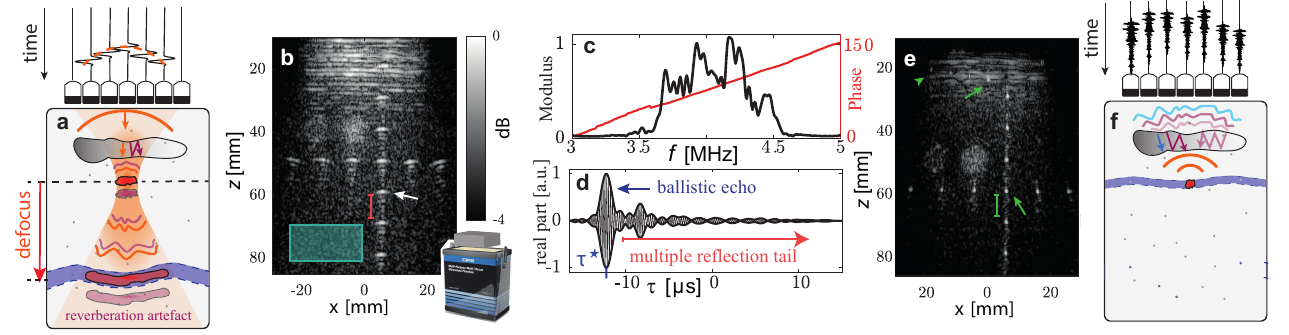}
    \caption{\textbf{Dispersive compensation of reverberations in speckle.} \textbf{a} Sketch of an incident focused wave-field in presence of a reverberating layer. The ballistic and multiply-reflected components focus at different depths shifted with respected to their respective isochronous volume. \textbf{b} Confocal image obtained for the homogeneous model $c_0 = 1540$ ms in the {phantom - Plexiglas} experiment. \textbf{c} Modulus and phase of the frequency dispersion law $\bar{H}(f)$ (Eq.~\ref{transfer}) extracted by IPR in the speckle on the blue area in panel b. \textbf{d} Normalized real part of the temporal dispersion law $H(\tau)$ (Eq.~\ref{psf}), highlighting a first peak at $\tau_{0,0}^{*} \sim -12.4$ $\mu$s~ associated with the ballistic echo and secondary peaks associated with multiple reflections. \textbf{e} Corrected confocal image $\mathcal{I}_1(\mathbf{r},\tau=0)$ obtained by applying a dispersive focusing law (Eqs.~\ref{correction} and \ref{I1}).  \textbf{f} Sketch of the dispersive focusing law tailored and applied in post-processing to make interfere constructively each multiply-reflected path with the ballistic component of the wave-field.}
    \label{ch4_IPR1D_abLaw}
 \end{figure}

We are primarily interested in the academic experiment described in Ref.~\onlinecite{Giraudat2025}, in which a plexiglas plate is inserted between the probe and the ultrasound phantom. The canonical reflection matrix $\mathbf{R}_{\mathbf{u}\bm{\theta}}(t)$ of the medium is acquired using the sequence described in {Table 1} of the same paper (Fig.~\ref{principle}a). The reverberating Plexiglas layer displays a much higher sound velocity ($c_r\sim 2750$ m.s$^{-1}$) than the phantom ($c = 1540$ m.s$^{-1}$). This academic experiment mimics a transcranial imaging configuration. The Plexiglass layer induces strong defocusing and reverberation artifacts (Fig.~\ref{ch4_IPR1D_abLaw}a) similar to the ones produced by a skull (Fig.~\ref{ch4_RMintro}d).

As described in the first paper of the series~\cite{Giraudat2025}, the time-focused de-scan reflection matrix $\mathbf{R}_{\bm{\Delta}\mathbf{r}}(\tau)$ is constructed by a delay-and-sum beamforming process applied to  $\mathbf{R}_{\mathbf{u}\bm{\theta}}(t)$, with an homogeneous velocity model corresponding to the speed-of-sound $c$ in the phantom (Fig.~\ref{principle}b). Since we are first restricting the reverberation compensation to a time correction of temporal dispersion of echoes, the study of the focused reflection matrix is restricted to the confocal signal: 
\begin{equation}
    \mathcal{I}_0(\mathbf{r}_{\textrm{in}},\tau)=R(\Delta \mathbf{r}=0,\mathbf{r}_{\textrm{in}},\tau)
\end{equation}
This confocal signal is the result of a standard beamforming process with identical focusing points at input and output ($\mathbf{r}_{\textrm{in}} =\mathbf{r}_{\textrm{out}}$, Fig.~\ref{principle}b) but letting the echo time evolve beyond the expected ballistic time. The confocal signal taken at time lapse origin $\tau=0$ corresponds to the raw confocal image displayed in Fig.~\ref{ch4_IPR1D_abLaw}b.
It exhibits an amplitude drop with depth. This attenuation is due to the decay of energy experienced by ultrasonic waves due to scattering and absorption. As shown in Appendix~\ref{A}, this phenomenon can be detrimental to the assessment of reverberations. A time gain compensation is thus applied to circumvent wave attenuation.

\subsection{Dispersive Phase Law }

A reverberating layer implies that each frequency component of the wave-field undergoes a different phase aberration. A temporal Fourier transform of the confocal signal is thus performed in order to proceed with a frequency analysis of the wave-field: 
\begin{equation}
    \overline{\mathcal{I}}_0\left(\mathbf{r},f\right) =  \int d \tau~    {\mathcal{I}}_0\left(\mathbf{r},\tau \right)  e^{-i 2 \pi f \tau}.
    \label{ch4_TFtemp}
 \end{equation} 
For lateral aberration correction~\cite{lambert_distortion_2020}, a broadband aberration phase law was extracted by considering the correlation between each spatial frequency component of the wave-fronts induced by virtual sources belonging to the same isoplanatic area. For axial aberration correction, a correlation between each temporal frequency component of the wave-fronts shall be considered instead. The difficulty here consists in finding an homogeneous speckle area that can provide the self-portrait of the wave-field. This area should be large enough in order to encompass a sufficiently high number of resolution cells to guarantee the convergence of the correlation matrix towards its covariance. However, it should avoid the presence of bright scatterers that would bias the estimation of a dispersive phase law. As a result, we here selected the area $\mathcal{A}$ of virtual sources depicted by the blue rectangle in Fig.~\ref{ch4_IPR1D_abLaw}b whose lateral and axial dimensions are $w_x = 22$ mm by $w_z = 10$ mm, respectively.

The correlation between each frequency component of the wave-field in the selected area is stored in a spectral correlation matrix $\mathbf{C}=[C(f,f')]$ whose coefficients are computed as follows:
\begin{equation}
   C(f,f') = \sum_{\mathbf{r} \in \mathcal{A}} \overline{\mathcal{I}}_0(\mathbf{r},f) \overline{\mathcal{I}}_0^*(\mathbf{r},f').
\end{equation}
As described in the first paper of the series, an iterative phase reversal process can be applied to the correlation matrix to synthesize a coherent guide star from each virtual source in the patch $\mathcal{A}$. The associated dispersive phase law is obtained using the following iterative process,
\begin{equation}
\label{ipr}
  { \mathbf{W}}^{\mathrm{(n+1)}} = \exp \left( i \arg \left \{  \mathbf{C} \times   {\mathbf{W}}^{\mathrm{(n)}} \right \}\right),
\end{equation}
with $ {\mathbf{W}}^{\mathrm{(0)}}$, an arbitrary wave-front that we will take arbitrarily as constant:  $ {\mathbf{W}}^{\mathrm{(0)}} = \left(1 \cdots 1\right)^\top$.
The phase $\bm{\phi}_H$ of the invariant obtained at the end of this process, $\exp \left ( i \bm{\phi}_H \right ) = \lim\limits_{\mathrm{n} \to \infty}  {\mathbf{W}}^{\mathrm{(n)}}$, is an estimator of the frequency dispersion law induced by the reverberating layer (Fig.~\ref{ch4_IPR1D_abLaw}c). As we will see further, this frequency dispersion law corresponds to the set of phase shifts to be applied to each of the frequency components of the echoes in order to make each multiple reflection path interfere constructively behind the reverberating plates. To check the consistency of this law over the frequency spectrum, a simple matrix product between $\mathbf{C}$ and ${\mathbf{W}}$ provides an estimator of the transfer function of the reverberating layer:
\begin{equation}
\label{transfer}
      \overline{\mathbf{H}} =  \mathbf{C} \times  \exp \left ( i \bm{\phi}_H \right ) .
\end{equation}
The amplitude of the resulting vector $\overline{\mathbf{H}}$ provides the frequency spectrum on which the IPR invariant is effective  (Fig.~\ref{ch4_IPR1D_abLaw}c). Contrary to a SVD process that would result into a quasi-monochromatic phase law~\cite{Giraudat2025}, the IPR process leads to a correction effective over the whole frequency bandwidth, here 3.5-4.5 MHz. The transfer function ${\overline{H}}(f)$ also displays an amplitude modulation that can be related to the Fabry-P\'{e}rot resonances supported by the reverberating layer (Fig.~\ref{ch4_IPR1D_abLaw}c).

The inverse Fourier transform of $\overline{\mathbf{H}}$ gives access to the temporal impulse response associated with the reverberating layers (Fig.~\ref{principle}c): 
\begin{equation}
\label{psf}
    {H}\left(\tau\right) =   \int df ~  \overline{{H}}\left(f\right) e^{i 2 \pi f \tau}.
\end{equation}
This temporal dependence of this impulse response ${H}\left(\tau\right)$  is displayed in Fig.~\ref{ch4_IPR1D_abLaw}d. It shows a main peak associated with the ballistic wave located at $\tau_{0,0}^\star \sim -12.4$ $\mu$s. This time shift of the ballistic wave is the signature of the mismatch between the wave velocity distribution and the velocity model $c$~\cite{Giraudat2025}. This impulse response also consists of secondary echoes of lesser intensity that can be attributed to multiple reflections induced by reverberating plate but also by the acoustic lens that lies at the surface of the probe~\cite{Giraudat2025}. They are the temporal counterparts of the Fabry-P\'{e}rot resonances highlighted by the transfer function $\overline{H}(f)$ displayed in Fig.~\ref{ch4_IPR1D_abLaw}c. 

The exact arrival times of these echoes observed in Fig.~\ref{ch4_IPR1D_abLaw}d can be determined analytically~\cite{Giraudat2025}. By considering the case of a double-layered medium, the iterative phase reversal process is shown to provide the reflected wave-field that would be obtained for a virtual reflector at the focusing point $\mathbf{r}^\star = (x,{z^\star}^{(0)})$ of the ballistic echo, with ${z^\star}^{(0)} =  z +  L (1 -  c_r/c)L$. The arrival time of each multiply-reflected reflected echo depends on the number of reflections undergone by the wave on the way out $(l)$ and on the way back $(m)$: 
\begin{equation}
    \tau^\star_{l,m}= {2\frac{L}{c}\left(\frac{c_r}{c} - \frac{c}{c_r}\right)} - \frac{2(l+m)L}{c_r}
\end{equation}
The temporal dispersion law here corresponds to the accumulated reverberations on the forward and return paths. It is not possible here to separate the temporal dispersion experienced by the incident beam from that experienced by the reflected wave. The expected reverberation time, $\tau_r=2L/c_r$ of the plexiglass plate ($c_r \sim 1000$ m.s$^{-1}$, $L=6$ mm) and of the acoustic lens is of 4.3 $\mu$s and 2$\mu$s. Hence, we probably observe a superimposition of multiply-reflected echoes induced by each layer in Fig.~\ref{ch4_IPR1D_abLaw}.

\subsection{Time Compression} \label{ch4_corr_globale_1D_f}
Because of the geometry of the reverberating plates and a wave velocity model $c_0$ matching with the phantom speed-of-sound, one can show that the aberration induced by the reverberation plate is spatially-invariant by translation.  Thus, as a first approximation, the frequency dispersion law extracted in the speckle area depicted in Fig.~\ref{ch4_IPR1D_abLaw}a can be considered as isoplanatic. This is why, a global image correction can be performed using this law. To do so, the phase conjugate of this law is applied to the confocal signal  $\overline{\mathcal{I}}_0(\mathbf{r},f)$ over the whole field-of-view (Fig.~\ref{principle}d), according to: 
\begin{equation}
\label{correction}
    \overline{\mathcal{I}}_1(\mathbf{r},f)=  \overline{\mathcal{I}}_0(\mathbf{r},f) \exp \left [ -i \phi_H (f) \right ].
 \end{equation}
 A corrected confocal signal, $\mathcal{I}_1(\mathbf{r},\tau)$, is then recovered by an inverse Fourier transform: 
 \begin{equation}
 \label{I1}
   \mathcal{I}_1(\mathbf{r},\tau) = \int  df ~   \overline{I}_1(\mathbf{r},f) e^{i 2\pi f d\tau}.
 \end{equation}
 Alternatively, the correction could have been performed in the time domain by convolving the reflection matrix with the time-reversed impulse response $H(\tau)$ (Fig.~\ref{principle}e). 

\begin{figure}\centering
    \includegraphics{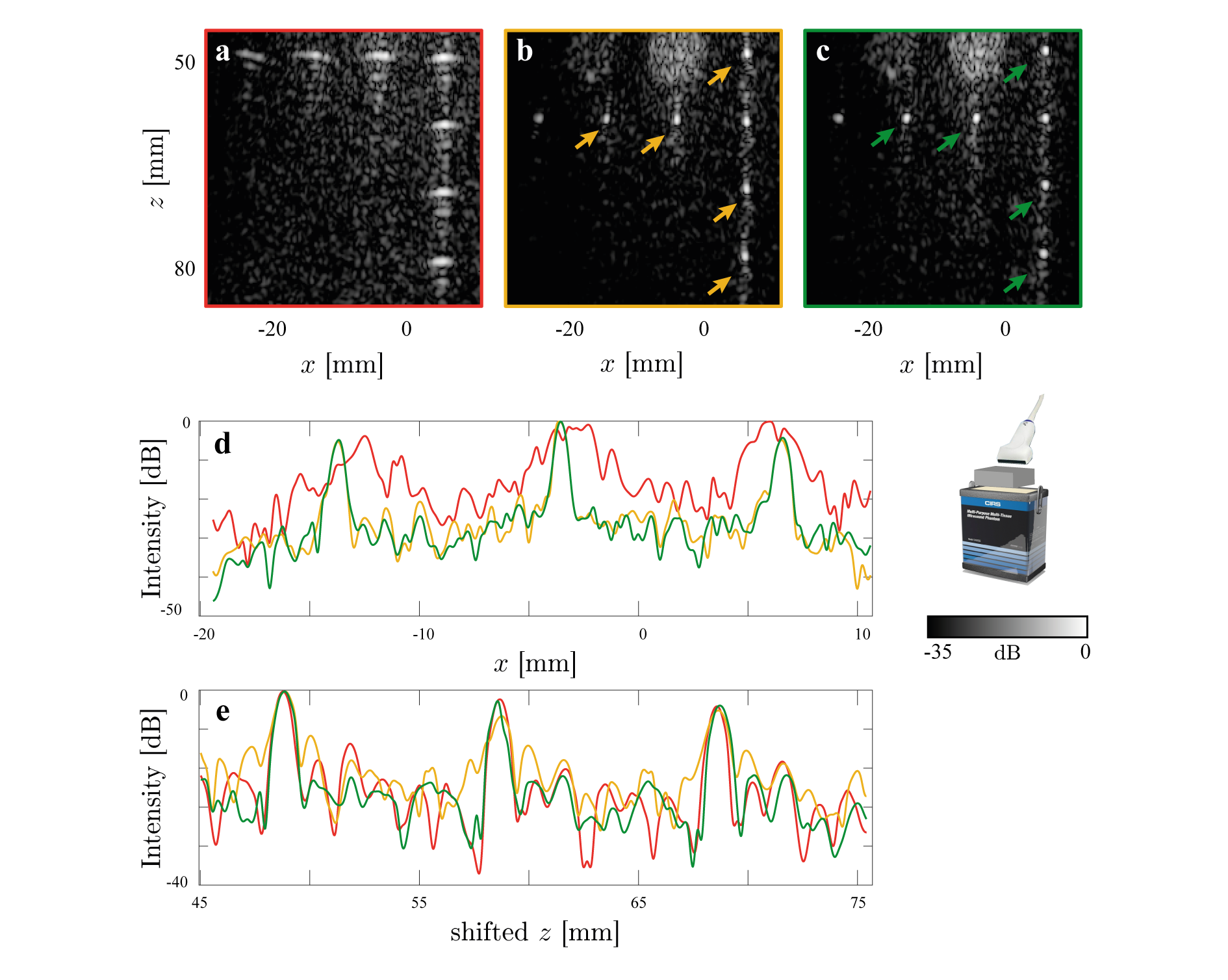}
    \caption{\textbf{Local spatio-temporal correction of aberrations in the phantom-plexiglass experiment.} \textbf{a} Confocal image $\mathcal{I}_0$ obtained for the homogeneous model $c_0 = 1540$ m.s$^{-1}$ in the {phantom - Plexiglas} experiment. \textbf{b} Image $\mathcal{I}_1$ (Eqs.~\ref{correction}-\ref{I1}) obtained after global frequency compensation of reverberations. \textbf{c} Image $\mathcal{I}_2$ (Eqs.~\ref{correct}-\ref{I2}) after spatio-temporal refocusing of ultrasound data. \textbf{d} Transverse cross-section of each image at the depth corresponding to the horizontal line of scatterers in the phantom. \textbf{e} Longitudinal cross-section of each image at the lateral position corresponding to the vertical line of scatterers in the phantom. In panels d and e, the original image $\mathcal{I}_0$ is displayed as a red continuous line while the corrected images $\mathcal{I}_1$ and $\mathcal{I}_2$ correspond to the yellow and green continuous lines, respectively.}
    \label{ch4_IPR2Dlocal_Results}
 \end{figure}

 The corrected confocal image, $ \mathcal{I}'(\mathbf{r},\tau=0)$, is shown in Fig.~\ref{ch4_IPR1D_abLaw}e. Its comparison with the initial image (Fig.~\ref{ch4_IPR1D_abLaw}b) shows the benefit of a dispersive phase correction. The targets now appear with much better contrast, comparable to that observed in a case without a reverberating layer~\cite{Giraudat2025}. The axial and transverse resolutions of their main lobe are also close to those obtained in an ideal case.  This improvement in image quality can be assessed quantitatively by plotting the transverse section of the images along lines of scatterers (Fig.~\ref{ch4_IPR2Dlocal_Results}d). It is made possible primarily by the time shift applied to the ballistic component, which adjusts the associated isochronous volume to its focusing depth $z^{\star{(0)}}$. It confirms that even when considering a purely temporal correction, we are able to act on the transverse spread of the focal spot when this one is mainly a defocus. The image improvement is also induced by a time shift of multiple reflections to make them interfere constructively with the ballistic echo at the lapse time $\tau=0$ (Fig.~\ref{ch4_IPR1D_abLaw}f).  
 
 However, we can see that this recombination is only partial and that axial secondary lobes associated with reverberations subsist and that their level is actually increased along the vertical line of scatterers (Fig.~\ref{ch4_IPR2Dlocal_Results}e). The reverberation plate not only induces an isoplanatic axial distortion of the wave-field. It also generates higher-order aberrations that requires a sharper tailoring of local spatio-temporal focusing laws. This is the aim of the next part.
 
\section{Spatio-Temporal Correction of Reverberations}\label{ch4_section_CorrSpatioFreq}
\begin{figure}[h!tb]\centering
    \includegraphics[width=\textwidth]{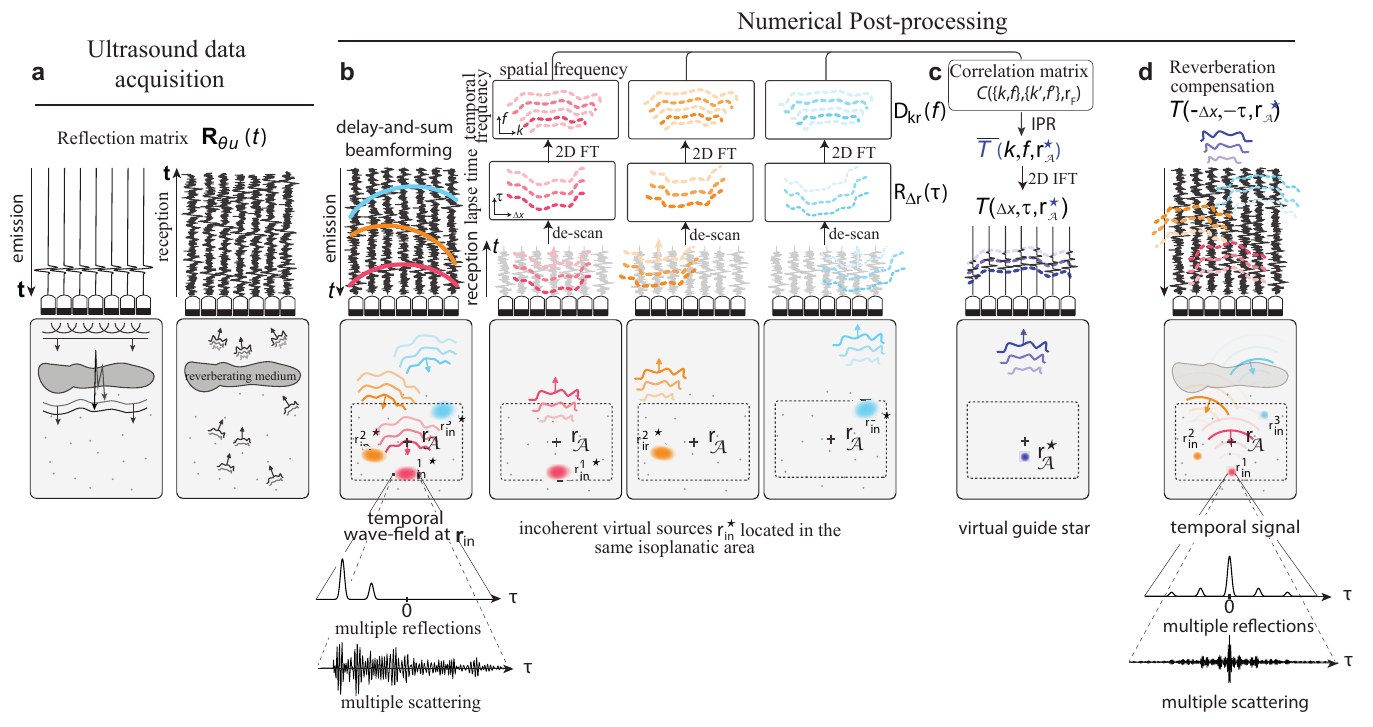}
    \caption{Extraction of a spatio-temporal focusing law in speckle. \textbf{a} Acquisition of the reflection matrix $\mathbf{R}_{\mathbf{u}\bm{\theta}}(t)$: (\textit{i}) each incident plane wave is multiply-reflected by the reverberating layer before being scattered by sub-resolved heterogeneities; (\textit{ii}) the reflected waves also undergo reverberations before being recorded by the ultrasonic probe. \textbf{b} A delay-and-sum beamforming process decoupling the input and output focal spots is applied to $\mathbf{R}_{\mathbf{u}\bm{\theta}}(t)$ to compute a time focused reflection matrix $\mathbf{R}_{xx}(z,\tau)$ at each depth $z$. A de-scan operation followed by a 2D Fourier transform yields the frequency-dependent distortion matrix $\mathbf{D}_{kx}(f)$ in the plane wave basis. \textbf{c} A correlation matrix is computed between the distorted wave-fields associated with each virtual source $\mathbf{r}_{\textrm{in}}$ belonging to the selected area $\mathcal{A}$ centered around the point $\mathbf{r}_{\mathcal{A}}$. An IPR analysis conducted at each spatio-temporal frequency leads to the synthesis of a coherent virtual source and the estimation of the reverberated wave-field $\overline{T}(k_x, f)$ in the Fourier space, or equivalently, to ${T}(\Delta x, \tau)$ in real space. \textbf{d} Time reversal of this spatio-temporal wave-field yields a focusing law that can compensate for reverberations, leading to a spatio-temporal recompression of the echoes associated with each point $\mathbf{r}$ of the medium.}
    \label{principle2}
 \end{figure}
Temporal and spatial aberration correction methods offer complementary approaches. In general, it may be necessary to combine them in order to shape a spatio-temporal focusing law adapted to the medium and correct the transverse and axial distortions of the image. 
 With this in mind, we now propose to combine the frequency correction method presented in Sec.~\ref{ch4_section_CorrFreq} with a transverse correction by generalizing the distortion matrix concept introduced in previous works~\cite{badon_distortion_2020,lambert_distortion_2020,bureau_three-dimensional_2023}.
 
 \subsection{Frequency Analysis of Wave Distortions}
 
 To implement this, the time focused reflection matrix~\cite{Giraudat2025} is now considered at each depth $z$ of the medium: $\mathbf{R}_{xx}(z,\tau) = [R\left(x_{\textrm{out}},x_{\textrm{in}},z,\tau \right)] $. It contains the cross-talk between virtual sources $\mathbf{r}_{\textrm{in}} = (x_{\textrm{in}},z)$ and the virtual detectors $\mathbf{r}_{\textrm{out}} = (x_{\textrm{out}},z)$ located at the same depth. In the first paper of the series~\cite{Giraudat2025}, the associated de-scan matrix, $\mathbf{R}_{\Delta x}(z,\tau)=[R(\Delta x,x_{\textrm{in}},z,\tau]$ (with $\Delta x=x_{\textrm{out}}-x_{\textrm{in}}$), allowed us to simultaneously probe the temporal dispersion and spatial spread of the focal spots. Here, we show how to harness its spatio-temporal degrees of freedom in order to compensate for the residual distortions exhibited by the confocal image in Fig.~\ref{ch4_IPR1D_abLaw}e.
 
To do so, a correction basis shall be first determined. Ideally, this basis should maximize the spatial correlations between the distorted wave-fields. In the present case, the lateral invariance of the reverberating layer implies a maximal isoplanicity of wave distortions in the plane wave basis. A spatial Fourier transform is therefore applied to the de-scan matrix $\mathbf{R}_{\Delta x}(z,f) $ (Fig.~\ref{principle2}b):
\begin{equation}
        \mathbf{D}_{kx}\left(z,f\right)=  \mathbf{F}^\top \times \overline{\mathbf{R}}_{\Delta x}\left(z,f\right)  ,
    \label{ch3_Rduale}
 \end{equation} 
 using the spatial Fourier transform operator $\mathbf{F}$: 
\begin{equation}
   F(x,k_x) = \exp\left(i k_x x\right),
\end{equation}
where $k_x$ represents the transverse spatial frequency of the associated plane wave. The result of Eq.~\ref{ch3_Rduale} is a distortion matrix whose columns contain the angular decomposition of the wave-front induced by each virtual source $\mathbf{r}_{\textrm{in}}=({x}_{\textrm{in}},z)$ (Fig.~\ref{principle2}b). It contains the signature of the wave-front distortions undergone by each reflected wave-field when going through the reverberating layer. 

\begin{figure}\centering
   \includegraphics{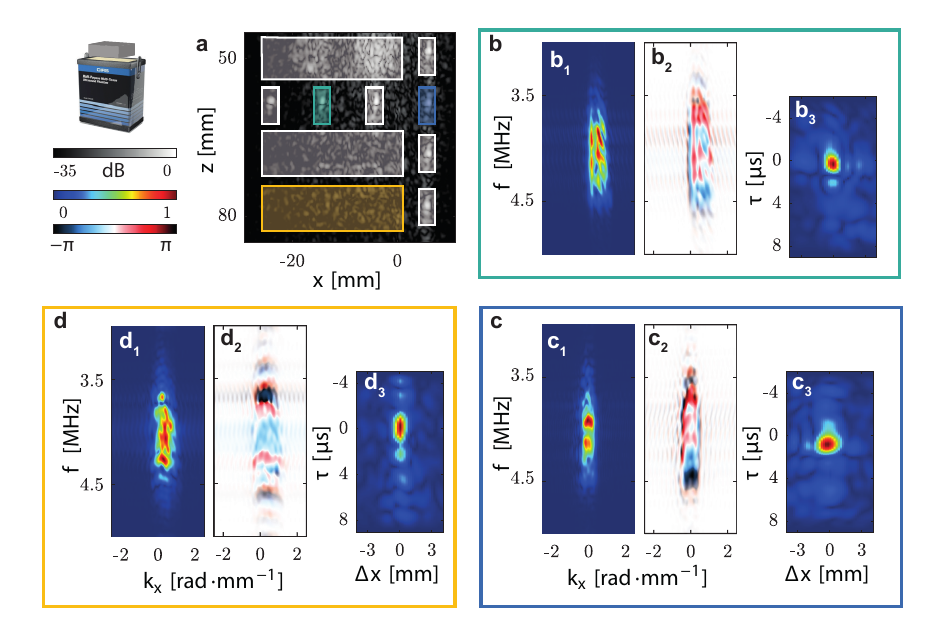}
   \caption{\textbf{Extraction of spatio-temporal focusing laws in the phantom through a reverberating layer.} \textbf{a} The field of view is divided into a set of areas for estimating the aberration law, as shown in white. \textbf{b}-\textbf{d} The laws obtained for the three colored areas are shown as examples. For each, from left to right: amplitude and phase of the law $\overline{\overline{T}}\left ( k_x, f \right)$, and amplitude of its spatio-temporal Fourier transform.}
   \label{ch4_IPR2Dlocal_ablaw}
\end{figure}

In previous works~\cite{lambert_distortion_2020,lambert_ultrasound_2022}, this distortion matrix was considered at the single ballistic time and its singular value decomposition was used to extract an aberration phase law at the central frequency, or equivalently, a set of time delays in the temporal domain. Here, the distortion matrix is frequency-dependent and we will now show how this polychromatic feature can be used to extract a spatio-temporal focusing law. To do so, we extend to the spatial frequency domain~\cite{bureau_three-dimensional_2023} the iterative phase reversal process only operated in the temporal frequency domain in the first part of the manuscript. To do so, a correlation matrix, $ \mathbf{C}_{\mathcal{A}}= \left [ C_{\mathcal{A}}(\lbrace k_x,f \rbrace ,\lbrace k'_x,f' \rbrace ) \right ]$, shall be built between each spatio-temporal frequency component $( k_x,f )$ of the distorted wave-fields over a given area $\mathcal{A}$ in the field-of-view:
\begin{equation}
     C_{\mathcal{A}}(\lbrace k_x,f \rbrace ,\lbrace k'_x,f' \rbrace ) = N_{\mathcal{A}}^{-1} \sum_{\mathbf{r}_{\textrm{in}}\in \mathcal{A}} D(k_x,f,\mathbf{r}_{\textrm{in}}) D^*(k'_x,f',\mathbf{r}_{\textrm{in}}) \exp \left \{- i \left [\phi_H(f)-\phi_H(f') \right ] \right \},
    \label{ch4_Ckf}
 \end{equation} 
where $N_{\mathcal{A}}$ is the number of resolution cells contained in ${\mathcal{A}}$. The choice of ${\mathcal{A}}$ is subject to the following dilemma. On the one hand, it should encompass a sufficiently large number of resolution cells in speckle to guarantee the convergence of $\mathbf{C}_{\mathcal{A}}$ towards its ensemble average, the covariance matrix $\langle \mathbf{C}_{\mathcal{A}} \rangle$. On the other hand, it should be as small as possible to grasp high-order aberrations associated with a small isoplanatic length. To this end, the field of view has been divided into a set of spatial windows $\mathcal{A}$ shown in Fig.~\ref{ch4_IPR2Dlocal_ablaw}a. Two types of areas can be distinguished: extended averaging areas in the speckle and more restricted areas on strong reflectors. Note that the phase term $\exp \left \{- i \left [\phi_H(f)-\phi_H(f') \right ] \right \}$ in Eq.~\ref{ch4_Ckf} enables a prior compensation of the frequency dispersion law determined at the previous step. This operation is made to facilitate the convergence of Eq.~\ref{ch4_Ckf} towards $\langle \mathbf{C}_{\mathcal{A}} \rangle$  by maximizing the correlation between each frequency component of the wave-field. 
%towards  We saw in the irst paper of the series that in the presence of an echogenic diffuser, RPI and SVD algorithms latch onto this bright diffuser rather than on a virtual reflector present at the focus [\cref{ch3_section_cible_c0bad}]. For this reason, in the first stage of frequency correction [\cref{ch4_section_CorrFreq}], we took care to avoid including strong reflectors in the averaging area in order to avoid any bias. Now that the difference between the focusing time and the ballistic time has been corrected across the entire field of view, the signal from strong diffusers can be easily used to compensate for aberrations.

 \subsection{Space-Time Transmittance}

The transmittance $\overline{\mathbf{{T}}}_{\mathcal{A}}=[{T}_{\mathcal{A}}(k_x,f)]$ of the reverberating layer seen from each point $\mathbf{r}$ in the field-of-view is extracted using the iterative phase reversal process (Eq.~\ref{ipr}) applied to the spatio-temporal frequency correlation matrix $\mathbf{C}_{\mathcal{A}}$ computed over the corresponding patch $\mathcal{A}$ (Eq.~\ref{ch4_Ckf}): 
\begin{equation}
    \overline{\overline{\mathbf{T}}}(\mathbf{r}\in \mathcal{A})= \mathbf{C}_{\mathcal{A}} \times  \lim\limits_{\mathrm{n} \to \infty}  {\mathbf{W}}^{\mathrm{(n)}}
\end{equation}
Figures~\ref{ch4_IPR2Dlocal_ablaw}b$_1$-d$_1$ and b$_2$-d$_2$ present the amplitude $|\overline{\overline{{T}}}_{\mathcal{A}}(k_x,f)|$ and phase $\phi_{{\mathcal{A}}}(k_x,f)$, respectively, of the transmittance $\overline{\overline{{T}}}_{\mathcal{A}}(k_x,f)$ obtained over the different colored areas depicted in Fig.~\ref{ch4_IPR2Dlocal_ablaw}a. 
%It can be noted that the time range considered has been further restricted around the ballistic time now that the defocus has been corrected and only the residual reverberations need to be compensated.
The associated focal spots, $T_{{\mathcal{A}}}(\Delta x, t)=\int df \int dk_x \overline{\overline{{T}}}_{{\mathcal{A}}}(k_x,f) e^{i k_x x} e^{-i 2\pi f t}$, are displayed in Figs.~\ref{ch4_IPR2Dlocal_ablaw}b$_3$-d$_3$. Figs.~\ref{ch4_IPR2Dlocal_ablaw}b$_3$ and d$_3$ display multiply-reflected echoes that were not perfectly corrected by the global dispersive law $\phi_H(f)$ at the first step of the process. The focal spot displayed in Fig.~\ref{ch4_IPR2Dlocal_ablaw}c$_3$ shows a time shift compared to the ballistic time and can be explained by the subsistence of a defocus after the global dispersion correction. The variations observed between these focal spots (Figs~\ref{ch4_IPR2Dlocal_ablaw}b$_3$-d$_3$) and the corresponding phase laws (Figs~\ref{ch4_IPR2Dlocal_ablaw}b$_2$-d$_2$) highlight the spatial fluctuations of the wave-distortions across the field-of-view. This justifies \textit{a posteriori} the need for a local estimation of the space-time transmittance.  The reverberation phase laws also show transverse phase fluctuations that demonstrate the interest of going beyond a dispersive focusing law and the need for a spatio-temporal correction of aberrations. As previously revealed by the global frequency response of the reverberating layer (Fig.~\ref{ch4_IPR1D_abLaw}c), the frequency spectra $|\overline{\overline{T}}(k_x,f)|$ of each aberration phase law (Figs.~\ref{ch4_IPR2Dlocal_ablaw}b$_1$-d$_1$) exhibit amplitude fluctuations that are reminiscent of the Fabry-P\'{e}rot resonances induced by the reverberating layer.

\subsection{Spatio-Temporal Focusing}

A phase conjugation of the transmittance matrix estimator at each frequency can be used to tailor spatio-temporal focusing laws. The compensation of reverberations can then be performed as follows:
\begin{equation}
\label{correct}
    R'(\Delta x, r_{\textrm{in}}\in \mathcal{A},\tau) =\int dk_x \int df  \exp \left ( -i [\phi_H (f) +\phi_{{\mathcal{A}}} (k_x,f)] \right ) R(k_x,r_{\textrm{in}},f) e^{ik_x \Delta x}e^{-i2\pi f \tau} 
    \end{equation}
This operation is therefore equivalent to a phase reversal experiment with both a recompression of the confocal signal both in time and space (Fig.~\ref{principle2}d). This is confirmed by the novel confocal image, $\mathcal{I}_2(\mathbf{r}_\textrm{in})$, that can be deduced from the confocal component ($\Delta x=0$) of the corrected de-scan matrix at the ballistic time ($\tau=0$):
\begin{equation}
\label{I2}
    \mathcal{I}_2(\mathbf{r}_{\textrm{in}})=R'(\Delta x=0, \mathbf{r}_{\textrm{in}},\tau=0)
\end{equation}
This result shown in Fig.~\ref{ch4_IPR2Dlocal_Results}c is compared with the initial image (Fig.~\ref{ch4_IPR2Dlocal_Results}a) and the intermediate one (Fig.~\ref{ch4_IPR2Dlocal_Results}b) obtained after application of a global dispersive focusing law at the first step of correction. The additional local space-time correction provides a better contrast and a better axial resolution through a sharper recombination of multiple reflections, as highlighted by the arrows pointing on bright scatterers in Fig.~\ref{ch4_IPR2Dlocal_Results}. To be more quantitative, Fig.~\ref{ch4_IPR2Dlocal_Results}d displays a transverse cross-section of the confocal image along the horizontal line of scatterers in the phantom. Compared to the initial image shown in red, the compensation of reverberations leads to a remarkable contrast enhancement of 10 dB for the point-like scatterers throughout the field-of-view. Fig.~\ref{ch4_IPR2Dlocal_Results}e displays the longitudinal cross-section of the confocal image along the vertical line of scatterers in the phantom. Compared to the dispersive correction that previously increased the level of secondary lobes along this line of scatterers (orange line in Fig.~\ref{ch4_IPR2Dlocal_Results}e), the spatio-temporal correction restores and even slighty improves the initial contrast while drastically improving the transverse resolution of the image. 

Nevertheless, we can see that the recombination of multiple reflections is only partial and that side lobe effects associated with reverberations remain in Fig.~\ref{ch4_IPR2Dlocal_Results}c. This is partly due to the phase-matched nature of the filter chosen for the correction~\cite{Herrin1977}. If we wanted to obtain a perfect image, we would need to be able to whiten the frequency spectrum of the image which is altered by the Fabry P\'{e}rot resonances of the reverberating plate (Fig.~\ref{ch4_IPR2Dlocal_ablaw}). This is the principle of an inverse filter but this operation would be extremely sensitive to noise and here hampered by out-of-focus echoes.

\section{Application to transcranial imaging}

Now that the distortion matrix concept has been extended to the frequency domain, the method is applied to the more realistic case of a head phantom. In a previous study~\cite{bureau_three-dimensional_2023}, matrix imaging has been used to determine phase shift laws that compensate for the aberrations suffered by direct echoes through the skull of the head phantom. Here we show how the approach developed above can compensate for the temporal dispersion induced by multiple reflections at the skull interfaces.

\subsection{Experiment on an Adult Head Phantom}
\begin{figure}\centering
    \includegraphics[width=\textwidth]{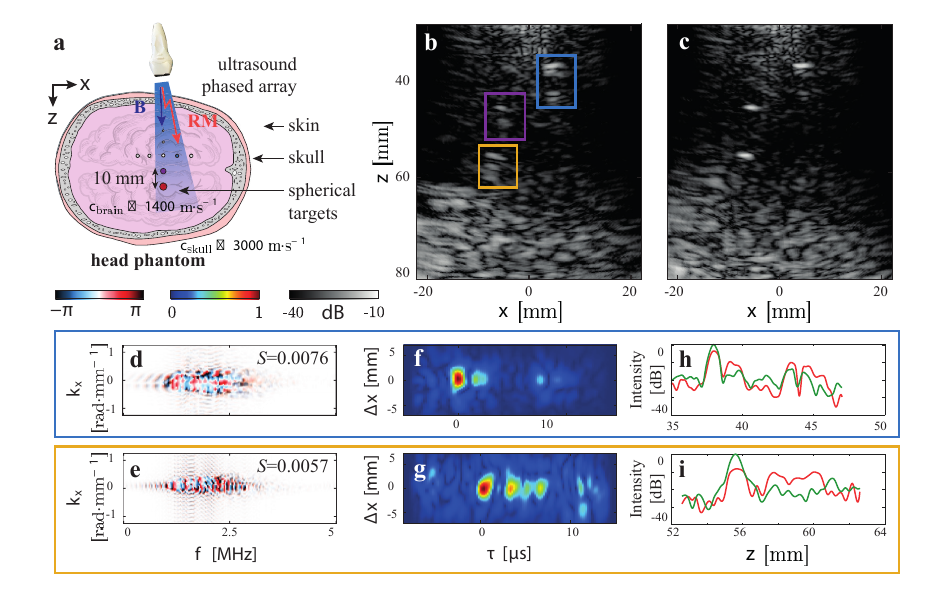}
    \caption{\textbf{Experiment on head phantom.} \textbf{a} A linear phased array probe is placed at the temporal window of the head phantom. The blue beam indicates the field of view captured by the probe. \textbf{b} Original confocal image, under a velocity model $c_0 = 1400$ m.s$^{-1}$. \textbf{c} Image obtained after local space-time correction using the signal from strong scatterers. \textbf{d}, \textbf{e} Phase distribution of the transmittance $\overline{\overline{T}}(k_x,f)$ estimated over the blue and yellow areas displayed in panel b. The amplitude of $\overline{\overline{T}}(k_x,f)$ is encoded with a transparency mask. \textbf{f}, \textbf{g} Associated spatio-temporal focal spots $T(\Delta x,\tau)$. \textbf{h}, \textbf{i} Depth dependence of the confocal signal $\mathcal{I}(\mathbf{r},\tau=0)$ at the target location before (red line) and after reverberation compensation (green line).}
    \label{ch4_HPsetup}
 \end{figure}

Spatio-temporal matrix imaging is now applied to an adult head ultrasound phantom. This phantom (True Phantom Solutions) allows us to reproduce \textit{in vitro} the conditions of brain imaging in adults  (Fig.~\ref{ch4_HPsetup}a). It consists of a material that mimics the acoustic characteristics of the human skull and an internal tissue that mimics those of biological soft tissues (Tab.~\ref{ch4_tableHP}). Highly reflective spherical inclusions made of bone-mimicking material are placed inside the phantom. They are evenly spaced at $10$ mm~ and arranged in a three-dimensional cross pattern. Their diameter increases with depth: $0.2$, $0.5$, $1$, $2$, and $3$ mm~(Fig.~\ref{ch4_HPsetup}a).

\begin{table}[h!tb]
    \centering
       \begin{tabular}{|c|c|c|c|} 
       \hline
       \textbf{phantom} & \textbf{speed-of-sound}  & \textbf{density}   & \textbf{attenuation}  \\
       \textbf{area} &[m.s$^{-1}$] & [\unit{\gram\per\cubic\centi\meter}] & at $2.25$ MHz~ [\unit{\dB\per\centi\meter}]\\
       \hline
       \hline
       cortical bone & $3000\pm30$  & $2.31$ & $6.4\pm0.3$ \\
       \hline
        trabecular bone & $2800\pm50$  & $2.03$ & $21\pm2$ \\
        \hline
        cerebral tissues & $1400\pm10$  & $0.99$ & $1.0\pm0.2$ \\
        \hline
        skin tissues & $1400\pm10$  & $1.01$ & $1.7\pm0.2$ \\
        \hline
    \end{tabular}
    \caption{\textbf{Acoustic properties of the head phantom.}}
    \label{ch4_tableHP}
 \end{table}

 The ultrasonic acquisition of the reflection matrix is performed as follows (see also Tab.~\ref{Table_SondePA_ExpSKULL}). A linear phased array probe is placed on the temporal window of the head phantom where the skull thickness is approximately $~6$ mm, as shown in Fig.~\ref{ch4_HPsetup}a.  Impedance matching was performed using ultrasonic contact gel. The reflection matrix was recorded in the plane wave basis to allow for a better signal-to-noise ratio~\cite{montaldo_coherent_2009}. A sequence of $81$ plane waves is emitted, assuming a sound velocity of $c_0 = 1400$ m.s$^{-1}$, in accordance with the expected velocity in the soft tissues of the phantom. The resulting confocal image is shown in Fig.~\ref{ch4_HPsetup}b. It suffers from significant aberrations, as highlighted by the three targets in the field of view, which appear with significant reverberation echo tails. 
\begin{table}[h!tb]
    \centering
       \begin{tabular}{|c|l|l|l|} 
       \hline
       \multicolumn{3}{|c|}{\textbf{parameter}} & \textbf{value} \\
       \hline
       \hline
       \multirow{5}*{\textbf{probe}} & \multicolumn{2}{l|}{type} & linear, phased array \\
       % \cline{3-4}
        &\multicolumn{2}{l|}{number of transducers} & $N_u = 96$ \\
        & \multicolumn{2}{l|}{inter-element distance}  & $\delta u = 0.2$~\unit{\milli\metre} $\sim  \lambda/2$\\
        &  \multicolumn{2}{l|}{aperture} & $\Delta u = 19.2$~\unit{\milli\metre} \\
        & \multicolumn{2}{l|}{central frequency}  & $f_c = 2.8$~\unit{\mega\hertz} \\
       \hline
       % \hline
       \multirow{7}*{\textbf{Acquisition}} & \multicolumn{2}{l|}{imaging platform} & Aixplorer\textregistered, Supersonic Imagine\\
 & \multicolumn{2}{l|}{speed-of-sound model} & $c_0 = 1400$~\unit{\metre\per\second }\\
      & \multirow{3}*{plane wave sequence} & maximal angle & $\theta_{\textrm{max}}= \ang{40}$ \\
        &  & angular pitch & $\delta\theta = \ang{1}$ \\
        &  & number & $N_{\theta} = 81$\\
        & \multicolumn{2}{l|}{emitted signal} & two half-period pulse at $f_c$  \\
        & \multicolumn{2}{l|}{sampling frequency} & $f_s = 5$~\unit{\mega\hertz}\\
        & \multicolumn{2}{l|}{recording time} & $\Delta t = 217$~\unit{\micro\second}\\   
        \hline
    \end{tabular}
    \caption{\textbf{Probe and acquisition sequence for recording the canonical reflection matrix $\R_{u\theta}(t)$ in the head phantom experiment.}}
    \label{Table_SondePA_ExpSKULL}
 \end{table}

 \subsection{Local Space-Time Correction of Skull Reverberations}

Following the approach described above, an iterative phase reversal process (Eq.~\ref{ipr}) is applied to the correlation matrix of wave distortions expressed in the $(k,f)$-space (Eq.~\ref{ch4_Ckf}). Transmittance phase distribution $\overline{\overline{T}}_{\mathcal{A}}(k_x,f)$ are extracted in the colored areas shown in Fig.~\ref{ch4_HPsetup}b and surrounding the bright scatterers. Even though the frequency bandwidth is limited to 1-3 MHz due to skull attenuation, the phase distribution $\phi_{\mathcal{A}}(k_x,f)$ of the estimated transmittance is particularly complex whether it be along the temporal or spatial frequency dimension. This complexity can be assessed by computing the Strehl ratio $\mathcal{S}$ that provides the ratio between the energy at focus with and without aberrations and reverberations:
\begin{equation}
\mathcal{S}=\frac{\left | \left \langle \overline{\overline{T}}(k_x,f) \right \rangle_{k_x,f} \right |^2 }{\left \langle \left | \overline{\overline{T}}(k_x,f)\right |^2 \right \rangle_{k_x,f} }
\end{equation}
where the symbol $\langle \cdots \rangle$ stands for an average over the subscripts, here the spatial and temporal frequencies $(k_x,f)$. The transmittances displayed in Figs.~\ref{ch4_HPsetup}d and e display Strehl ratios $\mathcal{S}$ of 7.6$\times 10^{-3}$ and 5.7$\times 10^{-3}$, respectively. These extremely low values illustrate the detrimental impact of skull heterogeneities on the focusing process. The corresponding focal spots $T_{\mathcal{A}}(\Delta x, \tau)$ are also displayed in Figs.~\ref{ch4_HPsetup}f and g. In each case, the focal spot displays a tail of multiply-reflected echoes lasting for more than 10 $\mu$s  (Figs.~\ref{ch4_HPsetup}f and g). This time dispersion is a manifestation of the strong reverberation phenomena taking place between the two skull interfaces. 

As before, the estimated reverberation phase laws (Figs.~\ref{ch4_HPsetup}d and e) are used to correct the confocal image (Eqs.~\ref{correct}-\ref{I2}). The resulting image is displayed in Fig.~\ref{ch4_HPsetup}c. These complex focusing laws compensate for most of the reverberations that initially polluted the confocal image (Fig.~\ref{ch4_HPsetup}b). The new confocal image thus constitutes a more accurate estimator of the brain phantom reflectivity. As shown by the confocal signal at the target location displayed in Fig.~\ref{ch4_HPsetup}h and i, the image of the bright scatterers shows a drastic gain of in terms of contrast ($\sim10$ dB at $z=56$ mm, see Fig.~\ref{ch4_HPsetup}i).

%Note that this image has been obtained by combining the the local spatio-temporal focusing laws estimated on each of the bright diffusers. The correction laws were applied only where they were estimated (at the diffusers). No corrections were applied to the rest of the field of view, due to the strong anisoplanetism observed between the estimated transmittances.
%\textcolor{red}{Elsa, peux tu expliquer comment tu as fait pour corriger l'image? Si les lois de focalisation n'ont pas ete ppliquées en dehors des trois diffuseurs, pourquoi l'image confocale semble t-elle changer en bas du champ de vision? }

 \subsection{Limit in the Speckle Regime}

Even though the difference between Figs.~\ref{ch4_HPsetup}b and \ref{ch4_HPsetup}g is spectacular, we have not been able to extract a spatio-temporal focusing law in any speckle area of the brain for several reasons:
 \begin{itemize}
    \item[$\bullet$] the extremely weak signal-to-noise ratio due to skull attenuation
    \item[$\bullet$] the limited size of the purely speckle areas not affected by bright scatterers in the space-time ranges $(\Delta x, \tau)$
    \item[$\bullet$] the strong anisoplanetism of transcranial aberrations due to the curved shape of the skull layer, its heterogeneous composition, and its irregular thickness. 
\end{itemize}
The latter aspect can be illustrated by comparing the phase distributions of the skull transmittance seen from two different locations inside the brain phantom (Figs.~\ref{ch4_HPsetup}d and e). The two transmittances actually exhibit a weak similarity index of 0.13. This limited isoplanicity is critical in speckle because it imposes a limit on the size of the correlation area $\mathcal{A}$. This is a fundamental problem because our transmittance estimator suffers from a bias $| \delta \overline{T}(k_x,f) |^2$ scaling as:
\begin{equation}
  \left  | \delta \overline{T}(k_x,f)  \right |^2 \propto \frac{N_{{T}}}{N_{\textrm{in}}}.
\end{equation}
where $N_{{T}}\sim \mathcal{S}^{-1}$ is the number of coherence grains exhibited by the transmittance $\overline{\overline{T}}$. As the reverberation laws are quite complex in the present case ({$N_{T}\sim 200$}), the required number $N_{\textrm{in}}$ of focusing points is way too large to meet an even partial condition of isoplanicity. This is why we were unable to observe convergence of our method towards a satisfying focusing law in speckle in the present case.

\section{Discussion}

Even though this last experiment has outlined some of its limits for trans-cranial imaging, the proposed method constitutes a breakthrough in the field of ultrasound imaging. Admittedly, it can be seen as the digital counterpart of the various adaptive focusing methods already developed in the past~\cite{flax_phase-aberration_1988,montaldo_time_2011}. However, to our knowledge, these approaches had never been applied to the compensation of reverberations. Furthermore, they were particularly difficult to implement in practice because they required the iterative time reversal process to be repeated many times on all the speckle grains in the area under consideration. On the contrary, matrix formalism allows for compensation of aberrations and reverberations in post-processing. This feature offers much greater flexibility in terms of correction basis. It also allows an iteration of the adaptive focusing process, which is particularly critical when trying to correct for such high-order aberrations as reverberations. In fact, the number of iterations required to converge on the focusing laws is no longer limited by their physical execution, and the adaptive focusing process can be repeated for any correction plane and focal point of interest.
 
With regards to the isoplanatic limit, a transition to 3D imaging could help us to address a larger number of speckle grains in the same isoplanatic area. A previous study on this head phantom has actually shown that isoplanicity is greater in the $(x,y)$ plane than in the $(x,z)$ plane \cite{bureau_three-dimensional_2023}. Moreover, brain imaging is widely performed nowadays with contrast agents (microbubbles) that are exploited to beat diffraction in ultrasound localization microscopy~\cite{errico_ultrafast_2015}. Interestingly, microbubbles can also be used as ultrasound guide stars for adaptive focusing. While current aberration correction methods have been restricted so far to the compensation of wave distortions undergone by the ballistic wave-front~\cite{robin_vivo_2023}, the frequency-dependent distortion matrix approach can be exploited for tailoring spatio-temporal focusing laws that would enable the constructive interference of multiply-reflected paths with the ballistic component of the wave-field. At last, speckle dynamics provided by blood flow can also be of interest to multiply the realization of disorder for speckle~\cite{osmanski_aberration_2012} and ensure the convergence of the iterative phase reversal process despite a limited isoplanatic area.

Beyond trans-cranial imaging, the extension of matrix imaging to the temporal degrees of freedom can be fruitful for the inspection of other organs in ultrasound in which reverberation is a long-standing problem. For instance, the abdominal layer induce strong multiple reflections that often prevent  imaging the liver at shallow depth~\cite{Ferraioli2024}. Beyond ultrasound, the proposed approach can be applied to other field of wave physics in which the multi-element technology allows a reflection matrix to be measured. In seismology, multiples are known to often pollute the migration process and have been tackled by Marchenko redatuming~\cite{Wapenaar2014} in multi-layered configurations. Matrix imaging can, nevertheless be a more robust and adaptive tool than the existing approaches that require a precise knowledge of the seismic wave velocity inside the medium~\cite{touma_distortion_2021,giraudat_unveiling_2023}. Complex areas such as volcanoes or fault zones are typical areas in which conventional migration or more sophisticated tools such as full waveform inversion have failed so far. Extending the distortion matrix concept to the frequency domain would lead to the design of spatio-temporal focusing laws that would allow a better monitoring of magma storage areas or fluid injection in depth. 

At an opposite scale, adaptive optics has been limited so far to a spatial correction of aberrations in microscopy~\cite{Booth2014}. Yet, light propagation in tissues is distorted by forward multiple scattering events~\cite{Dunsby2003}. A frequency-dependent analysis would be thus extremely rewarding for a proper realignment of these complex trajectories. While such a process is illusory experimentally with conventional adaptive optics, matrix imaging offers an adequate framework to perform this task in post-processing~\cite{kang_tracing_2023,najar_non-invasive_2023}. Several studies have actually shown how a multi-wavelength and multi-angle illumination scheme coupled to an interferometric measurement of the reflected wave-field can lead to a measurement of a multi-spectral distortion matrix~\cite{balondrade_multi-spectral_2023,zhang_deep_2023,lee_exploiting_nodate}. As demonstrated in this study, this quantity is actually the ideal starting point for the design of spatio-temporal focusing laws able to control the interference of multiple scattering paths in depth. 

To conclude, in this work, we have tackled the issue of reverberations in ultrasound imaging. The aim of this work was to propose a correction method that broadens the range of aberrations that can be compensated for. The spatial and temporal degrees of freedom offered by broadband transducer arrays have been exploited using a matrix formalism to probe the spatial and temporal dispersion of ultrasonic echoes and then to find the focusing law adapted to the medium heterogeneities. Extending the distortion matrix concept to the frequency domain and taking full advantage of the iterative phase reversal process in the $(f,k)$-space, the method was successfully applied to compensate for defocus and multiple reflections induced by a reverberating layer on a tissue-mimicking phantom in an academic experiment. The interest of this method has then been explored for trans-cranial imaging through an experiment on a head phantom. While the approach has shown its ability for designing adaptive spatio-temporal focusing laws in presence of bright scatterers, it has also shown some limits in speckle areas due to the strong heterogeneities of the skull that limit the range of the memory effect. Nevertheless, the proposed method paves the way towards numerous applications ranging from ultrasound to optical microscopy or seismology in which deep imaging is long-standing goal and reverberations a fundamental problem.

\clearpage 
\noindent\textbf{Acknowledgments.}
The authors are grateful for the funding provided by the European Research Council (ERC) under the European Union's Horizon 2020 research and innovation program (grant agreement 819261, REMINISCENCE project, AA). W.L. acknowledges financial support from the SuperSonic Imagine company.  \\

\noindent\textbf{Author contributions.}
A.A. and M.F. initiated the project. A.A. supervised the project. F.B. and W.L. performed the phantom experiment. E.G. developed the post-processing tools. E.G. and A.A. performed the theoretical analysis.  E.G. and A.A. prepared the manuscript. E.G., F.B., W.L., M.F. and A.A. discussed the results and contributed to finalizing the manuscript.

\appendix

\section{Time gain compensation}
\label{A}
\begin{figure}[h!tb]\centering
   \includegraphics{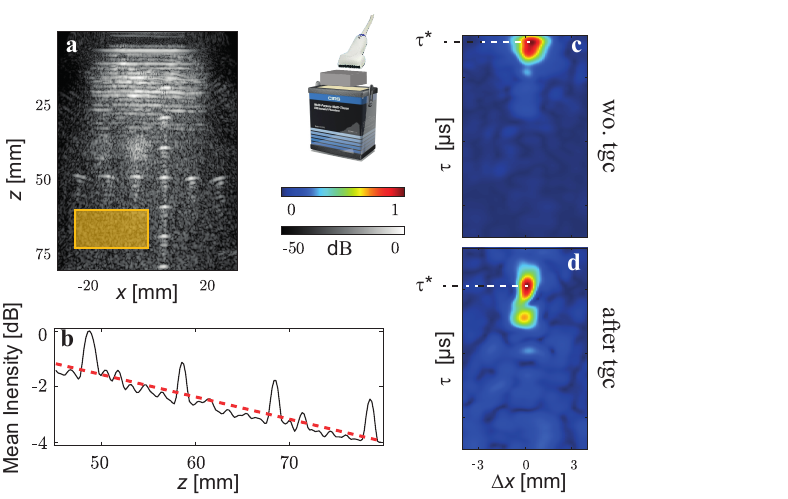}
  \caption{\textbf{Wave attenuation compensation.} \textbf{a} Raw confocal image. \textbf{b} Linear fit of the natural logarithm of the average confocal intensity (besides bright scatterers) as a function of depth leading to an estimation of the attenuation coefficient $\beta$. \textbf{c},\textbf{c} Estimated impulse response $H(\Delta x,\tau)$ obtained in the speckle area before (c) and after (d) attenuation compensation, respectively.}
   \label{tgc}
 \end{figure} 
In a statistically homogeneous disordered medium, the mean confocal intensity, $ \left \langle \left | \mathcal{I}_0(x,z) \right |^2 \right \rangle $, shall scale as $\exp (-2z /\ell_{\textrm{ext}})$, with $\ell_{\textrm{ext}}$, the extinction length. To retrieve such an exponential decay, the speckle-like fluctuations of the confocal image (Fig.~\ref{tgc}a) should be priorly smoothed out by lateral averaging. The resulting mean confocal intensity is displayed in log-scale as a function of effective depth $z$ (Fig.~\ref{tgc}b). Its evolution in speckle (\textit{i.e} outside of bright scatterers) can be fitted by an exponential curve (red dashed line in Fig.~\ref{tgc}b) whose decay provides an estimation of $\ell_{\textrm{ext}}\simeq 117$ mm. The overall fitting curve can be used to normalize at each time the raw data such that:
\begin{equation}
\label{tgc_com}
\mathbf{R}'_{u\theta}(t)=\mathbf{R}_{u\theta}(t)\exp \left [c_0 t /(2\ell_{\textrm{ext}}) \right ]
\end{equation}
Figure~\ref{tgc}c and d compares the impulse response $H(\Delta x, \tau)$ associated with the reverberating layer without and with the prior time gain compensation of Eq.~\ref{tgc_com}. In absence of time gain compensation, the amplitude of echoes decreases with echo time. It gives rise to a strong bias on our estimation of the impulse response (Fig.~\ref{tgc}c) with a shift towards negative time lapse of the ballistic echo time $\tau_{0,0}^{\star}$ and the underestimation of multiply-reflected echoes that arrive at larger times-of-flight. After time gain compensation (Fig.~\ref{tgc}d), the direct echo is retrieved at the expected ballistic time and multiple reflections are satisfactorily grasped by the estimated impulse response.  

\vspace{5 mm}

\bibliography{references2}

\end{document}